\documentclass[12pt,thmsa]{article}
\usepackage{amssymb}
\usepackage{graphicx}

\input{tcilatex}
\begin{document}

\begin{center}
\textbf{Dynamics of impurity, local and non-local information for two non
identical qubits}

$^{\dagger }${\small N. Metwally, M. Abdel-Aty}$^{\dagger }${\small \ and
M.Sebawe Abdalla}$\ddagger $

{\footnotesize $^{\dagger }$Mathematics Department, College of Science,
Bahrain University, 32038 Bahrain \\[0pt]
$^{\ddagger }$Mathematics Department, College of Science, King Saud
University, P. O. Box 2455, Riyadh 11451, Saudi Arabia}
\end{center}

From the separability point of view the problem of two atoms interact with a
single cavity mode is investigated. The density matrix is calculated and
used to discuss the entanglement and to examine the dynamics of the local
and non-local information. Our examination concentrated on the variation in
the mean photon number and the ratio of the coupling parameters.
Furthermore, we have also assumed that the atomic system is initially in the
ground states as well as in the intermediate states. It has been shown that
the local information is transferred to non-local information when the
impurity of one qubit or both is maximum.\newline

\bigskip


\bigskip

\section{Introduction}

As well known quantum entanglement is a non-local correlation in quantum
system which plays a fundamental role in almost all efficient protocols of
quantum computation and quantum information processing. This is due to the
fact that the quantum states of two or more coupled objects have to be
described with reference to each other, even though the individual objects
may be spatially well separated. It is widely recognized that the control of
quantum entanglement leads to new classes of measurement, communication and
computational systems, and in some cases dramatically to perform the
non-quantum analogies.

The research on quantifying entangled states has been considered by several
authors \cite{1,2,3}. To quantify entanglement we have to know whether the
states are pure or mixed states. Therefore, if the entangled state is in a
pure state, then it is sufficient to use von Neumann entropy \cite{4}, which
has a unique measurement. Here we may refer to the quantum computation of
trapped ions introduced by Cirac and Zoller \cite{5}. It can be regarded as
a potentially powerful technique for storage and manipulation of quantum
information. In this case the information is usually stored in the spin
states of an array of trapped ions and manipulated using laser pulses.
Reasonably long coherence times can be achieved, compared to achievable
switching rates \cite{6}, where individual qubits are addressed through
spatial separation of the ions.

Experimental implementations of this scheme have succeeded in performing
simple two-qubit logic gates \cite{7,8} and preparing entangled states \cite%
{9}. Recently, there has been an ongoing effort to characterize
qualitatively and quantitatively the entanglement properties of condensed
matter systems and apply them in quantum communication and information. For
example, an important emerging field is the quantum entanglement in solid
state system such as spin chains which are the natural candidates for the
realization of the entanglement compared with the other physics system. Also
the problem of atom-field interaction or trapped ions interacting with laser
light were suggested as one promising candidate to build a small scale
quantum computer. In fact these type of problems have attracted considerable
attention due to its potential application in high-resolution spectroscopy,
as well as the high-precision atomic fountain clock and the high-precision
spin polarization measurements.

The main purpose of the present work is to discuss the degree of
entanglement for the above quantum system. Therefore we have to calculate
the dynamical operators from which we are able to reach our goal. This can
be achieved either by find the solution of the equations of motion in the
Heisenberg picture or to employ the wave function in the Schr\"{o}dinger
picture to derive the unitary operator. In this context the later method
will be adopted and this can be seen in the following section. Section
\textbf{III} is devoted to discuss Peres's and degree of entanglement. This
is followed by section \textbf{VI} where the impurity as well as the
dynamics of information is considered. Finally, in section \textbf{V} our
conclusion is given.

\section{The Model}

During the last decade many theoretical and experimental efforts have been
done in order to study processes involving atoms inside a cavity, stimulated
by the experimental realization of a multi-photon micromaser \cite{suc97}.
In the rotating wave approximation, the interaction of the cavity mode with
the injected atoms is described by the Hamiltonian
\begin{equation}
\hat{H}=\hslash \omega \hat{a}^{\dagger }\hat{a}+\hslash
\sum_{i=1}^{2}\left( \frac{\omega ^{(i)}}{2}\hat{\sigma}_{z}^{(i)}+\lambda
_{i}(\hat{\sigma}_{-}^{(i)}\hat{a}^{\dagger }+\hat{\sigma}_{+}^{(i)}\hat{a}%
)\right) .  \label{5}
\end{equation}%
Our target of this section is to derive the time-dependent density matrix
which enables us to discuss some of the statistical properties of the
present model. This can be reached from the solution of the Schr\"{o}dinger
picture. For this reason let us assume that the initial state of the two
atoms are prepared in a superposition state which can be written as
\begin{equation}
|\psi _{12}(0)\rangle =a_{1}a_{2}|e,e\rangle +a_{1}b_{2}|e,g\rangle
+b_{1}a_{2}|g,e\rangle +b_{1}b_{2}|g,g\rangle ,  \label{6}
\end{equation}%
where $|\psi _{12}(0)\rangle =|\psi _{1}(0)\rangle |\psi _{2}(0)\rangle $
and we have defined
\begin{equation}
|\psi _{i}(0)\rangle =a_{i}|e\rangle +b_{i}|g\rangle ,\qquad i=1,2.
\label{7}
\end{equation}%
Also we have considered the field to be initially in the coherent state i.e
\begin{equation}
|\psi _{f}(0)\rangle =\sum_{n=0}^{\infty }q_{n}|n\rangle ,\qquad q_{n}=\frac{%
\alpha ^{n}}{\sqrt{n!}}\exp (-\frac{1}{2}|\alpha |^{2}).  \label{8}
\end{equation}%
Now we can write the time evolution of the wave function in the form
\begin{equation}
|\psi (t)\rangle =\mathcal{U(}t\mathcal{)}|\psi _{12}(0)\rangle \otimes
|\psi _{f}(0)\rangle   \label{9}
\end{equation}%
where $\mathcal{U}(t)=\exp \left( -iHt/\hslash \right) $ is the
time-dependent unitary operator. Since the invariant sub-space of the global
system can be considered as a set of complete basis of the atom-field,
therefore one can expand the wave function for the present system to take
the form
\begin{equation}
|\psi (t)\rangle =\sum_{n=0}^{\infty }q_{n}\left[ A_{n}(t)|e,e,n\rangle
+B_{n}(t)|e,g,n+1\rangle +C_{n}(t)|g,e,n+1\rangle +D_{n}(t)|g,g,n+2\right]
\label{10}
\end{equation}%
where the coefficients $A_{n}(t),B_{n}(t),C_{n}(t)$ and $D_{n}(t)$ are given
by%
\begin{equation}
\left(
\begin{array}{c}
A_{n}(t) \\
B_{n}(t) \\
C_{n}(t) \\
D_{n}(t)%
\end{array}%
\right) =\left(
\begin{array}{cccc}
\mathcal{U}_{11}(t) & \mathcal{U}_{12}(t) & \mathcal{U}_{13}(t) & \mathcal{U}%
_{14}(t) \\
\mathcal{U}_{21}(t) & \mathcal{U}_{22}(t) & \mathcal{U}_{23}(t) & \mathcal{U}%
_{24}(t) \\
\mathcal{U}_{31}(t) & \mathcal{U}_{32}(t) & \mathcal{U}_{33}(t) & \mathcal{U}%
_{34}(t) \\
\mathcal{U}_{41}(t) & \mathcal{U}_{42}(t) & \mathcal{U}_{43}(t) & \mathcal{U}%
_{44}(t)%
\end{array}%
\right) \left(
\begin{array}{c}
a_{1}a_{2} \\
a_{1}b_{2} \\
b_{1}a_{2} \\
b_{1}b_{2}%
\end{array}%
\right)   \label{11}
\end{equation}%
and the entities $\mathcal{U}_{ij}(t)$ are calculated to take the form%
\begin{eqnarray}
\mathcal{U}_{11}(t) &=&-\frac{1}{(\mu _{n}-\nu _{n})}\left[ \beta
_{n}^{2}(1+r^{2})[\cos (\sqrt{\mu _{n}}t)-\cos (\sqrt{\nu _{n}}t)]+[\mu
_{n}\cos (\sqrt{\mu _{n}}t)+\nu _{n}\cos (\sqrt{\nu _{n}}t)]\right] ,
\nonumber \\
\mathcal{U}_{12}(t) &=&-\frac{ir}{(\mu _{n}-\nu _{n})}\left[ (\Delta
_{n}\beta _{n}+\gamma _{n}\mu _{n})\frac{\sin (\sqrt{\mu _{n}}t)}{\sqrt{\mu
_{n}}}+(\Delta _{n}\beta _{n}-\gamma _{n}\nu _{n})\frac{\sin (\sqrt{\nu _{n}}%
t)}{\sqrt{\nu _{n}}}\right] ,  \nonumber \\
\mathcal{U}_{13}(t) &=&\frac{i}{(\mu _{n}-\nu _{n})}\left[ (\Delta _{n}\beta
_{n}-\gamma _{n}\mu _{n})\frac{\sin (\sqrt{\mu _{n}}t)}{\sqrt{\mu _{n}}}%
-(\Delta _{n}\beta _{n}-\gamma _{n}\nu _{n})\frac{\sin (\sqrt{\nu _{n}}t)}{%
\sqrt{\nu _{n}}}\right] ,  \nonumber \\
\mathcal{U}_{14}(t) &=&-\frac{2r\Delta _{n}}{(\mu _{n}-\nu _{n})(1-r^{2})}%
\left[ \cos (\sqrt{\mu _{n}}t)-\cos (\sqrt{\nu _{n}}t)\right] ,  \nonumber \\
\mathcal{U}_{22}(t) &=&-\frac{1}{(\mu _{n}-\nu _{n})}\left[ (r^{2}\beta
_{n}^{2}+\gamma _{n}^{2}-\mu _{n})\cos (\sqrt{\mu _{n}}t)-(r^{2}\beta
_{n}^{2}+\gamma _{n}^{2}-\nu _{n})\cos (\sqrt{\nu _{n}}t)\right] ,  \nonumber
\\
\mathcal{U}_{23}(t) &=&\frac{r\delta _{n}}{(\mu _{n}-\nu _{n})(1+r^{2})}%
\left[ \cos (\sqrt{\mu _{n}}t)-\cos (\sqrt{\nu _{n}}t)\right] ,  \nonumber \\
\mathcal{U}_{24}(t) &=&\frac{i}{(\mu _{n}-\nu _{n})}\left[ (\Delta
_{n}\gamma _{n}-\beta _{n}\mu _{n})\frac{\sin (\sqrt{\mu _{n}}t)}{\sqrt{\mu
_{n}}}-(\Delta _{n}\gamma _{n}-\beta _{n}\nu _{n})\frac{\sin (\sqrt{\nu _{n}}%
t)}{\sqrt{\nu _{n}}}\right] ,  \nonumber \\
\mathcal{U}_{33}(t) &=&-\frac{1}{(\mu _{n}-\nu _{n})}\left[ (r^{2}\gamma
_{n}^{2}+\beta _{n}^{2}-\mu _{n})\cos (\sqrt{\mu _{n}}t)-(r^{2}\gamma
_{n}^{2}+\beta _{n}^{2}-\nu _{n})\cos (\sqrt{\nu _{n}}t)\right] ,  \nonumber
\\
\mathcal{U}_{34}(t) &=&-\frac{ir}{(\mu _{n}-\nu _{n})}\left[ (\Delta
_{n}\gamma _{n}+\beta _{n}\mu _{n})\frac{\sin (\sqrt{\mu _{n}}t)}{\sqrt{\mu
_{n}}}+(\Delta _{n}\gamma _{n}-\beta _{n}\nu _{n})\frac{\sin (\sqrt{\nu _{n}}%
t)}{\sqrt{\nu _{n}}}\right] ,  \nonumber \\
\mathcal{U}_{44}(t) &=&-\frac{1}{(\mu _{n}-\nu _{n})}\left[ \gamma
_{n}^{2}(1+r^{2})[\cos (\sqrt{\mu _{n}}t)-\cos (\sqrt{\nu _{n}}t)]+[\mu
_{n}\cos (\sqrt{\mu _{n}}t)+\nu _{n}\cos (\sqrt{\nu _{n}}t)]\right] ,
\nonumber \\
&&  \label{12}
\end{eqnarray}

The other components of the unitary operator can be deduced from the
relation $\mathcal{U}_{ij}(t)=$ $\mathcal{U}_{ji}(t),$ for $i\neq j=1,2,3,4.$
In the above equation we have used the abbreviations%
\begin{eqnarray}
\mu _{n} &=&\frac{1}{2}\left( \delta _{n}+\sqrt{\delta _{n}^{2}-4\Delta
_{n}^{2}}\right) ,\qquad \nu _{n}=\frac{1}{2}\left( \delta _{n}-\sqrt{\delta
_{n}^{2}-4\Delta _{n}^{2}}\right) ,  \nonumber \\
\delta _{n} &=&\sqrt{2n+3}(1+r^{2}),\qquad \Delta _{n}=\sqrt{(n+1)(n+2)}%
(1-r^{2}),  \label{13}
\end{eqnarray}%
and defined $\gamma _{n}=\sqrt{n+1},$ $\beta _{n}=\sqrt{n+2}$ while we set $%
r=\lambda _{2}/\lambda _{1}$.

Having obtained the explicate time-dependent unitary operator we are
therefore in position to derive the density operator for the field or for
the atomic system. In fact this can be achieved if we trace out either the
field or the atom. Since we are interested in this communication to discuss
the purity as well as the local and non-local information for the atomic
system. Therefore we have to trace over the field to obtain the density
matrix for the atom where $\rho _{12}=tr_{f}\{\rho _{s}\},$ and $\rho
_{s}(t)=|\psi (t)\rangle \langle \psi (t)|.$

In what follows we turn our attention to employ the results obtained to
discuss some statistical properties for the present system. This will be
seen in the forthcoming sections.

\section{Peres's criterion and degree of entanglement}

In this section, we discuss the entanglement between the two atoms. As we
know the atoms in the atomic system are called separable (uncorrelated) if
all the spectrum of the partial transpose of the atomic system is
non-negative. Therefore, it will be more convenient to use the
Peres-Horodecki's criterion to study the spectrum behavior of the partial
transpose of the state $\rho _{12}(t)$ \cite{18}$.$ In the meantime, to
quantify the amount of entanglement in the correlated states, one may use
the measurement introduced by K.Zyczkowski \cite{19}. Here, the degree of
entanglement ($DOE$) will be measured according to
\begin{equation}
DOE:=\sum_{i=1}^{N}|\eta _{i}|-1,  \label{15}
\end{equation}%
\begin{figure}[tbp]
\begin{center}
\includegraphics[width=18pc,height=12pc]{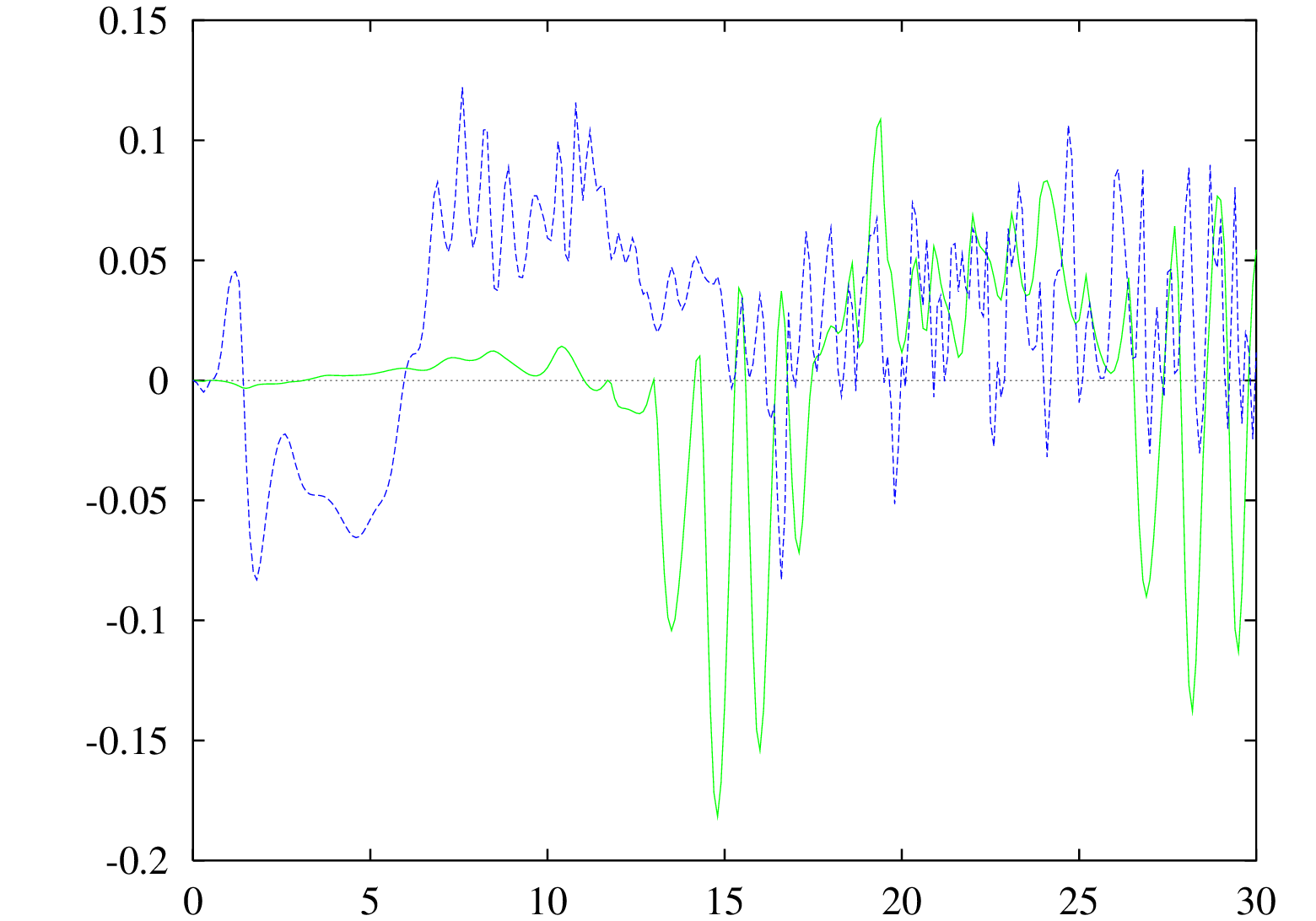}\ \put(-25,15){(a)}\
\put(-119,-10){ Scaled time} \put(0,75){PPT} %
\includegraphics[width=18pc,height=12pc]{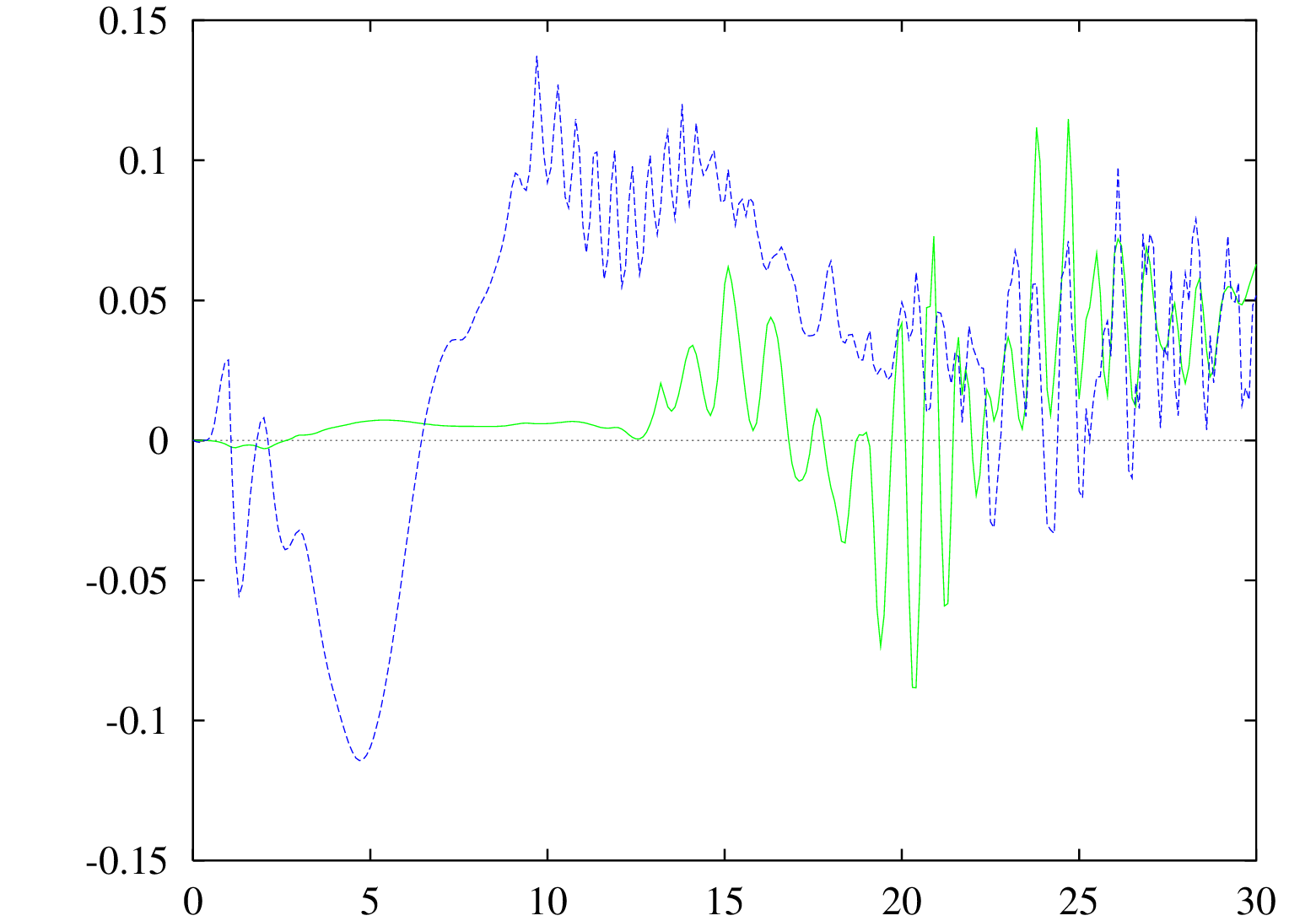}\ \put(-25,15){(b)}\
\put(-119,-10){ Scaled time} \put(-435,80){PPT}
\end{center}
\caption{ The Positive partial transpose criterion, PPT for atomic system is
prepared initially in the ground state $|gg\rangle $ for a coupling
operator, $r=0.1$( dash line) and $r=0.8$(solid curve)(a) For $\bar{n}=5$
(b) For $\bar{n}=10$.}
\end{figure}
where $\eta _{i},(i=1,2,3,...N)$ denotes the eigenvalues of the partial
transposed matrix $\rho ^{T}.$\ Note that for any separable matrix all
eigenvalues are positive and its trace is unity and hence $DOE$ is equal to
\emph{zero}. On the other hand, the maximally entangled states is \emph{one}
for a system belonging to a $2\times 2$ where the spectrum of eigenvalues $%
\eta _{i}$ consists of \{$-\frac{1}{2},\frac{1}{2},\frac{1}{2},\frac{1}{2}$%
\}, which in fact is consistent with the present system. Therefore, the
above equation is adequate for our case to measure the degree of
entanglement, however, we have to use a computational program due to the
complicated expression for the density matrix. To do so we have plotted some
figures to see the effect of the ratio of the coupling parameter $r=\lambda
_{2}/\lambda _{1}$ and the mean photon number $\bar{n}$ on the separability
as well as on the degree of entanglement.

The behavior of the positive partial transpose, PPT is plotted in Fig(1)
against the scaled time $\lambda _{1}t$. In Fig. \textrm{(1a)} and for a
fixed value of the mean photon number $\bar{n}=5$, we considered the cases
in which $r=0.1$ (dash line) and $0.8$ (solid line). In these two cases we
have assumed that the atomic system is initially in the ground state. For $%
r=0.1$ the entanglement occurs after onset of the interaction, however, for
short periods of time. This is followed by a period of disentanglement
between the atoms. The second period of entanglement has been observed,
however, for longer time compared with the first period. Moreover, the
entanglement in this case is apparent which means that as the period of time
increases the entanglement gets pronounced. This conclusion is limited for a
certain period of time where the atoms get separable again. Furthermore, the
irregular fluctuations which can be seen in the function behavior indicates
that the energy response to the correlation between the atoms gets weak.
This leads to a long instability and the separability behavior can be
reported. As time goes on the entanglement between the atoms is seen for a
long time. The situation is different when we increase the value of the
coupling parameter $r=0.8.$ This is realized from the solid curve where a
small amount of entanglement can be seen just for a short period of time.
This is followed by a long period of separability between the atoms. The
entanglement starts again to be seen at later time with amount greater than
that of the case in which $r=0.1$, but for short periods of time. This means
that the amount of entanglement is effected by the value of the coupling
parameter. Similar behavior can be seen for the case in which the mean
photon number $\bar{n}=10,$ however, slight different observation can be
observed. For example, for $r=0.1$ we observe an increase in the amount of
entanglement occurred in the second period of time compared with the case of
$\bar{n}=5$. While for $r=0.8$ the period of entanglement is shifted with
decreasing in its amount, see Fig. \textrm{(1b)}. This behavior is due to
the usual effect of the Rabi oscillations where the photon number plays a
crucial role in its value.

Now let us discuss the entanglement when the atomic system is initially
prepared in the partially entangled state such that
\begin{equation}
|\psi _{12}(0)\rangle =\cos \theta |e,e\rangle +\sin \theta |g,g\rangle ,
\label{16}
\end{equation}

\begin{figure}[tbp]
\begin{center}
\includegraphics[width=18pc,height=12pc]{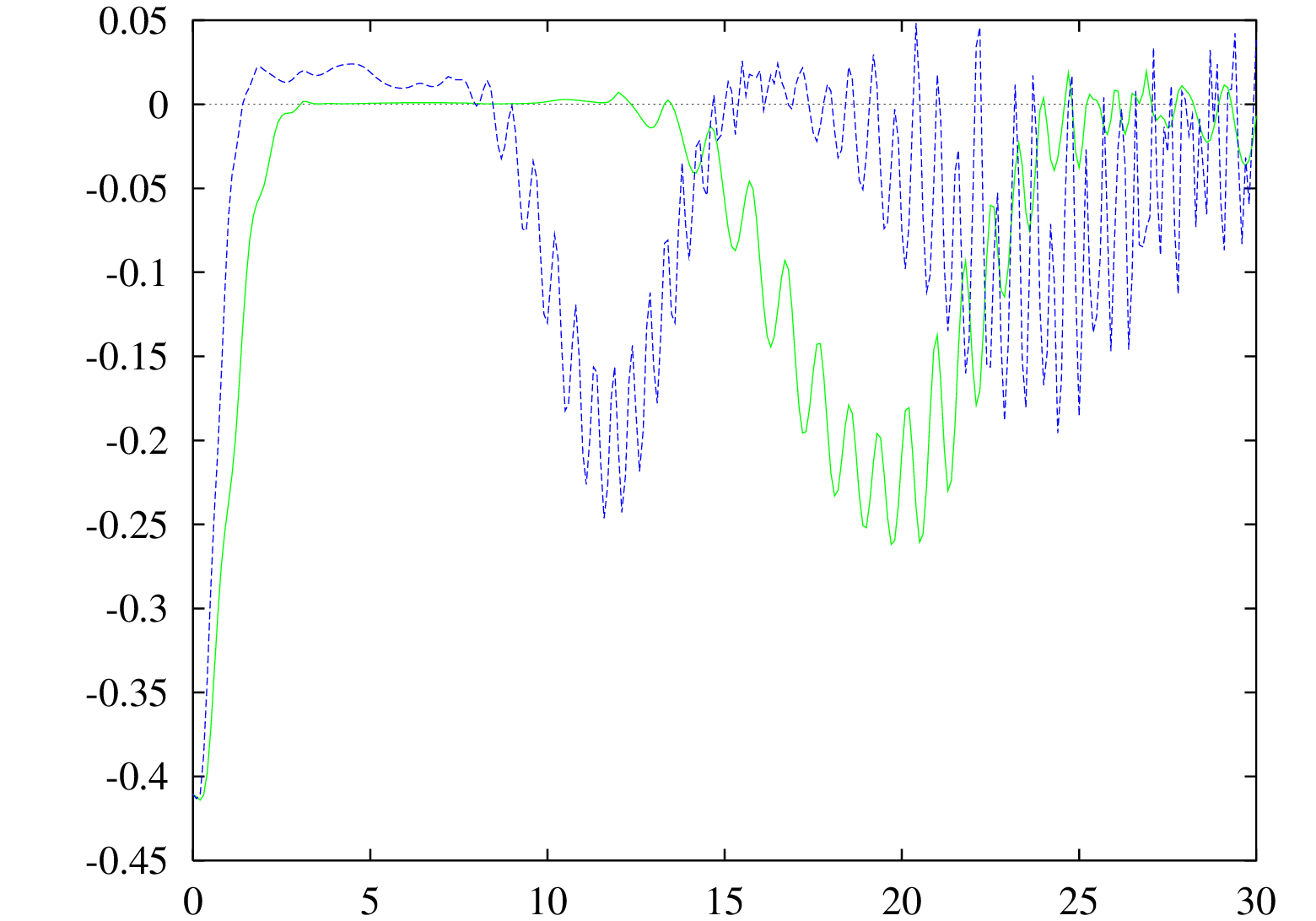}\ \put(-25,15){(a)}\
\put(-119,-10){ Scaled time} \put(0,80){PPT} %
\includegraphics[width=18pc,height=12pc]{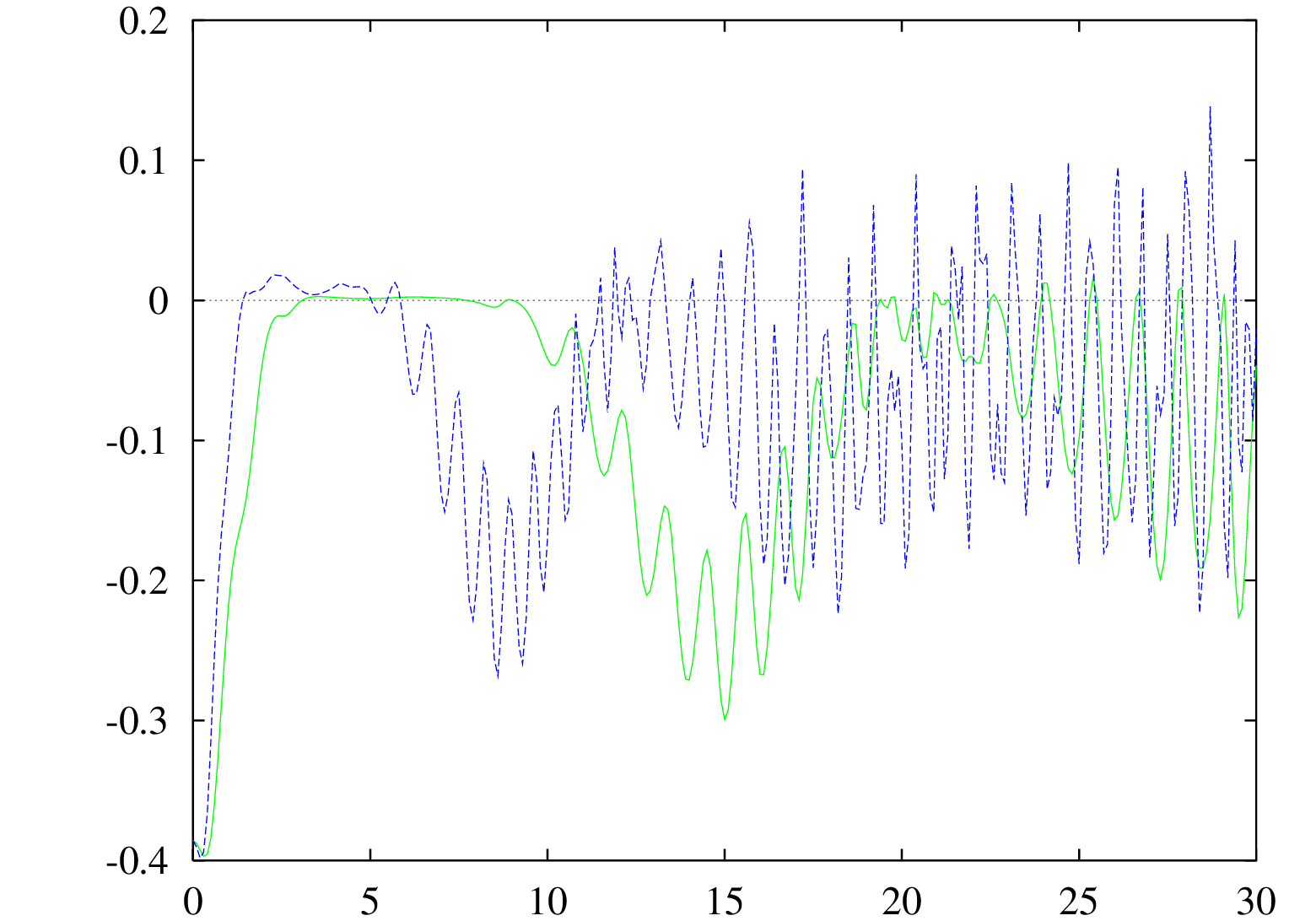}\ \put(-25,15){(b)}\
\put(-119,-10){ Scaled time} \put(-440,80){PPT}
\end{center}
\caption{ The same as Fig.$(1)$, but the system is The PPT criterion for
atomic system is prepared initially in a partial entangled state.}
\end{figure}
where $\theta =\pi /3.$ In this case and for $r=0.1,0.8$ and $\bar{n}=5$. In
this case there are different features have been seen. For example, strong
entanglement is realized for both cases with maximum value at $\sim -0.42$
occurred after onset of the interaction. This is followed by decreasing in
the degree of entanglement for both cases too, which means that the atoms
start to be separable. For the case $r=0.1$ the atoms get separable faster
than the case of $r=0.8,$ however the function shows irregular fluctuations
in the period of separability.

On the other hand, for the case in which $r=0.8$ more stability can be
realized for the disentanglement period. Also the second period of
entanglement is occurred for $r=0.1$ faster than the case when $r=0.8$.
However, with maximum value of the entanglement is less than that observed
of the first period of correlation. Furthermore, the general behavior shows
irregular fluctuations which referees to the energy exchange between the
atoms and the field, see Fig.\textrm{(2a)}. Similar behavior can be reported
for the case in which $\bar{n}=10$, but with more fluctuations for the case
in which $r=0.1,$ see Fig.\textrm{(2b)}.
\begin{figure}[tbp]
\begin{center}
\includegraphics[width=20pc,height=12pc]{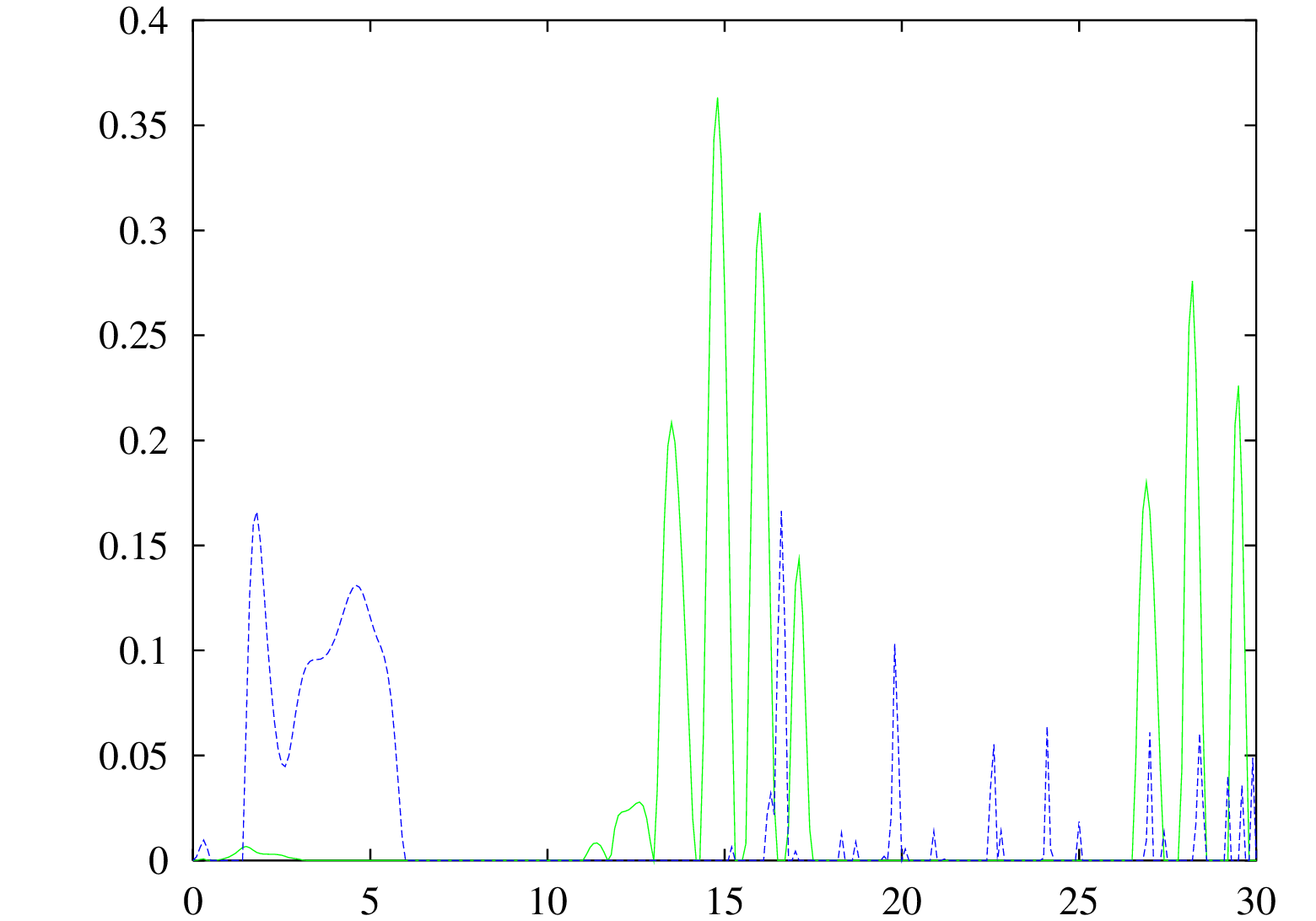}\ \put(-26,112){(a)}\
\put(-125,-10){ Scaled time}\put(-8,70){DOE} %
\includegraphics[width=20pc,height=12pc]{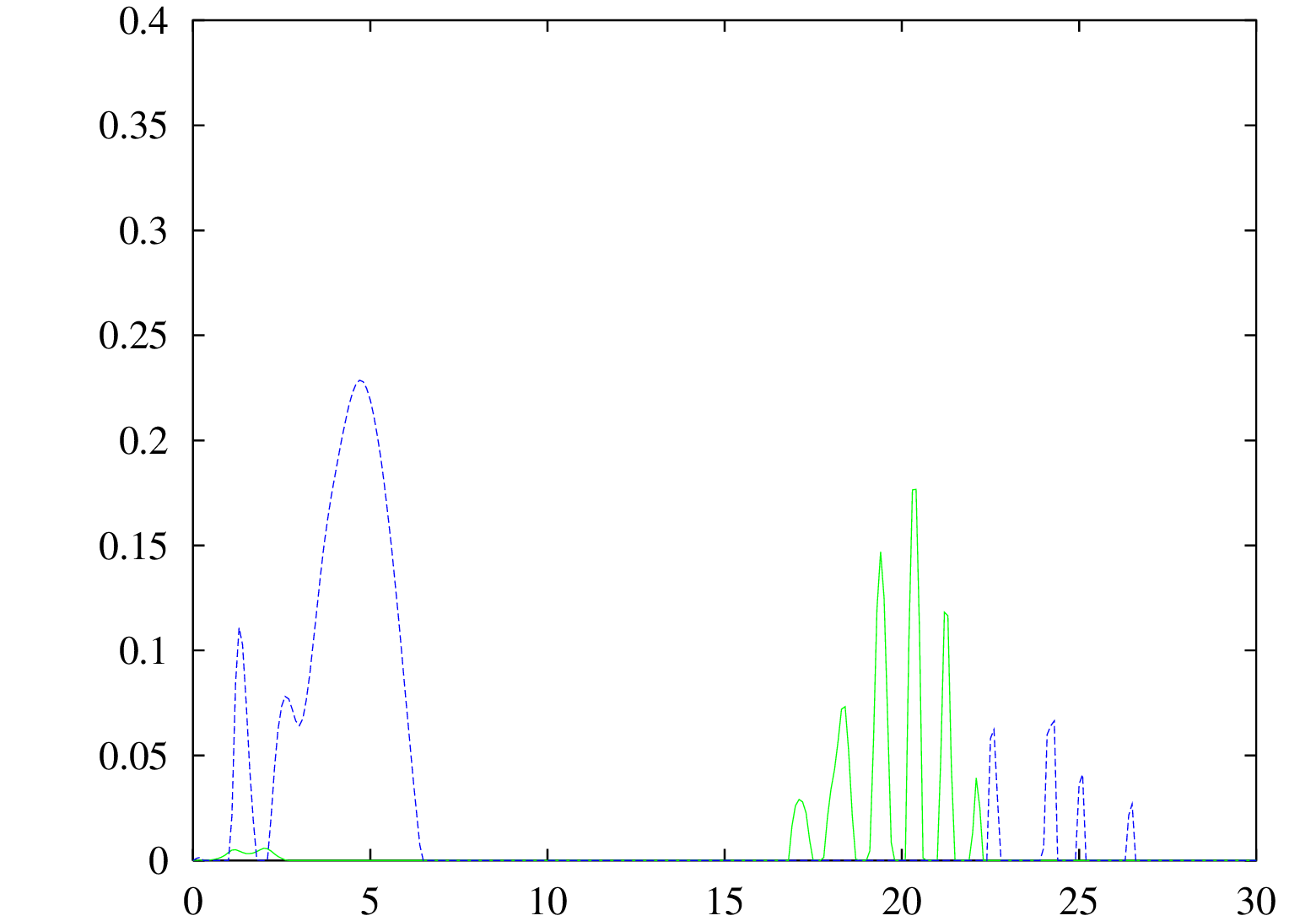}\ \put(-26,112){(b)}\
\put(-125,-10){ Scaled time}\put(-490,70){DOE}\ %
\includegraphics[width=25pc,height=10pc]{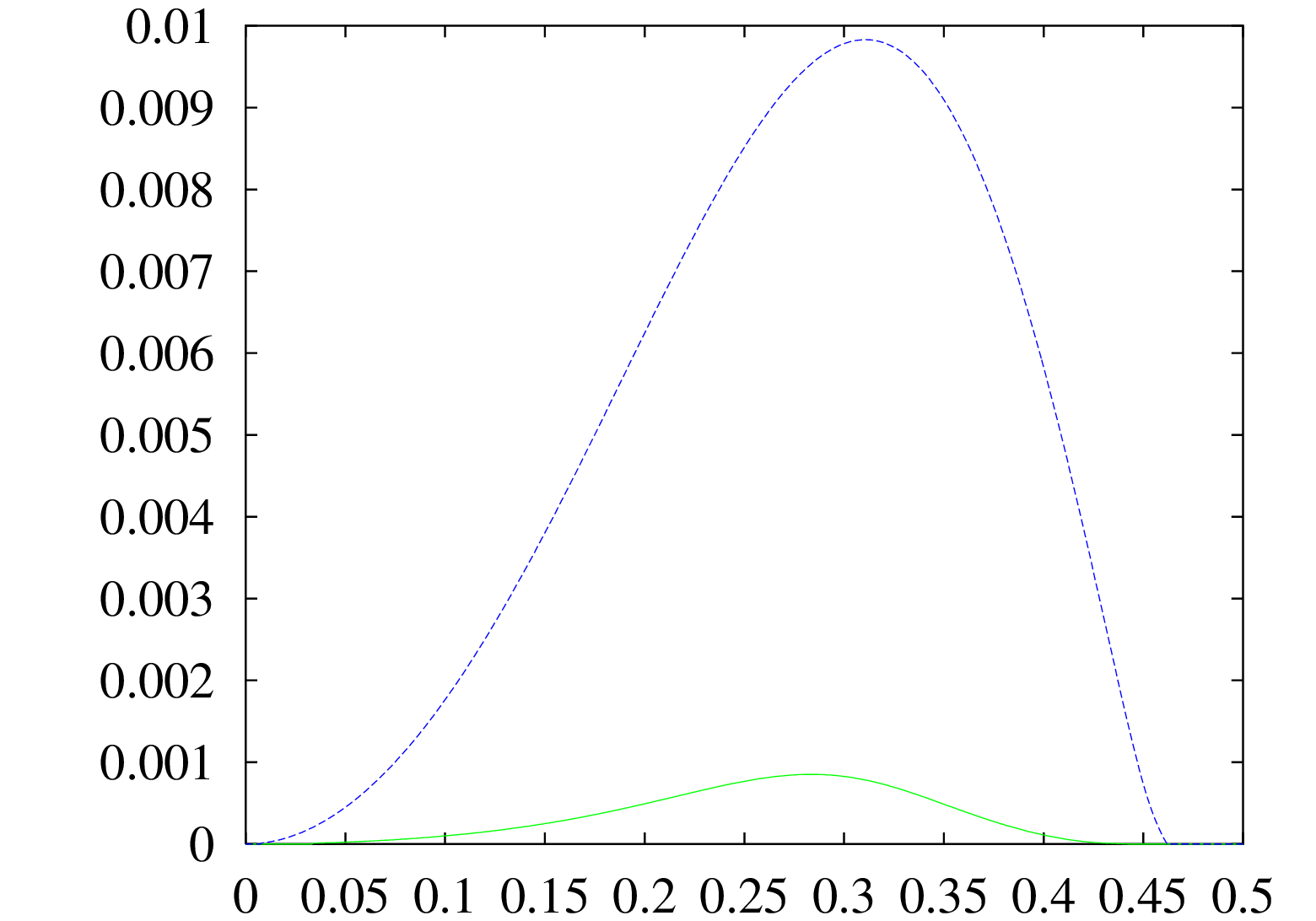}\ \put(-160,-10) { Scaled
time}\put(-300,65){\small{DOE}} \put(-35,100){(c)}
\end{center}
\caption{ The degree of entanglement, DOE for atomic system is prepared
initially in the ground state for $R=0.1$( solid curve line) and $R=0.8$(dot
curve)(a) For $\bar{n}=5$ and (b) for $\bar{n}=10$ (c) The same as
Fig.(a)but for small values of the scaled time.}
\end{figure}

To quantify the amount of entanglement contained in the atomic system we
have plotted figures \textrm{(3)}. The system is considered to be initially
in the ground state using the same value of the parameters as before. For
example, in Fig.\textrm{(3a)} we have considered $r=0.1$ (dash line) and $%
r=0.8$ (solid line) and the mean photon number $\bar{n}=5.$ In this figure
the system shows weak entanglement for short period of time after onset of
the interaction. This is followed by short period of time where the atoms
are separable and consequently the amount of entanglement is zero. This
means that the energy of the system is totally transferred to the field.
However, strong correlation between the atoms can be seen in the second
period of entanglement compared with the first period. In the meantime, the
maximum value of entanglement in this case approaches $0.17.$ Also, we
observe a slight changes of fluctuations in this period which reflects the
exchange of the energy between the two atoms. In fact this is the longest
period of entanglement during the considered time and also it is the most
pronounced one. The third period of correlation is short with a small amount
of entanglement and occurred after long period of time. This is followed by
the fourth period of correlation (short but with maximum value $\sim 0.17$)
and then the system starts showing a successive periods of entanglement with
decreasing in their amounts, see Fig.3a. Increasing the value of the
coupling parameter ratio $r=0.8,$ different observation can be reported
where a small amount of entanglement can be seen after a short period of
time.

\bigskip

On contrary to the previous case the strong entanglement occurred after
considerable period of time with maximum value approaches $0.36$. Also a
regular fluctuations with different maximum values can be seen, which means
that the exchange of the energy between the atoms and the field is nearly
stable. This behavior should be expected since the values of the coupling
parameters get close to each other. We have also examined the case in which $%
\bar{n}=10,$ for $r=0.1$ and $r=0.8.$ Increasing the value of the mean
photon number would effect on the Rabi frequency and hence on the system
behavior. This can be realized from the figure \textrm{(3b)} where the
amount of entanglement for $r=0.1$ increased compared with the case $\bar{n}%
=5.\ $In the meantime, for the case $r=0.8$, we can see decreasing in the
amount of entanglement which occurred after considerable period of time.

In Fig.(3c), we plot the degree entanglement for small scale of the scaled
time, where we consider the system is initially prepared in the ground
state. Since we start from a product state, the degree of entanglement is
zero at $\lambda_1 t=0$. As the time increases the degree of entanglement
increases. For large values of the coupling constant $r=0.8$, the degree of
entanglement is increased faster and reaches its maximum value(0.009), round
the scaled time $\lambda_1 t\simeq 0.3$. Then the sudden death phenomenon of
entanglement is seen for $\lambda_1 t>0.3$ \cite{Eberly}, where the degree
of entanglement is decreased quickly until completely vanishes. On the other
hand for small values of $r=0.1$, the degree of entanglement is slightly
increased reaching to its maximum value $( \simeq 0.001$), which is very
small comparing by the previous case. Then the degree of entanglement decays
smoothly till completely vanishes. This type of decay is seen for behavior
of the two qubit pair in Bloch channels \cite{nasser}.

Now let us turn our attention to consider the system to be initially in the
intermediate state, equation (15).
\begin{figure}[tbp]
\begin{center}
\includegraphics[width=20pc,height=12pc]{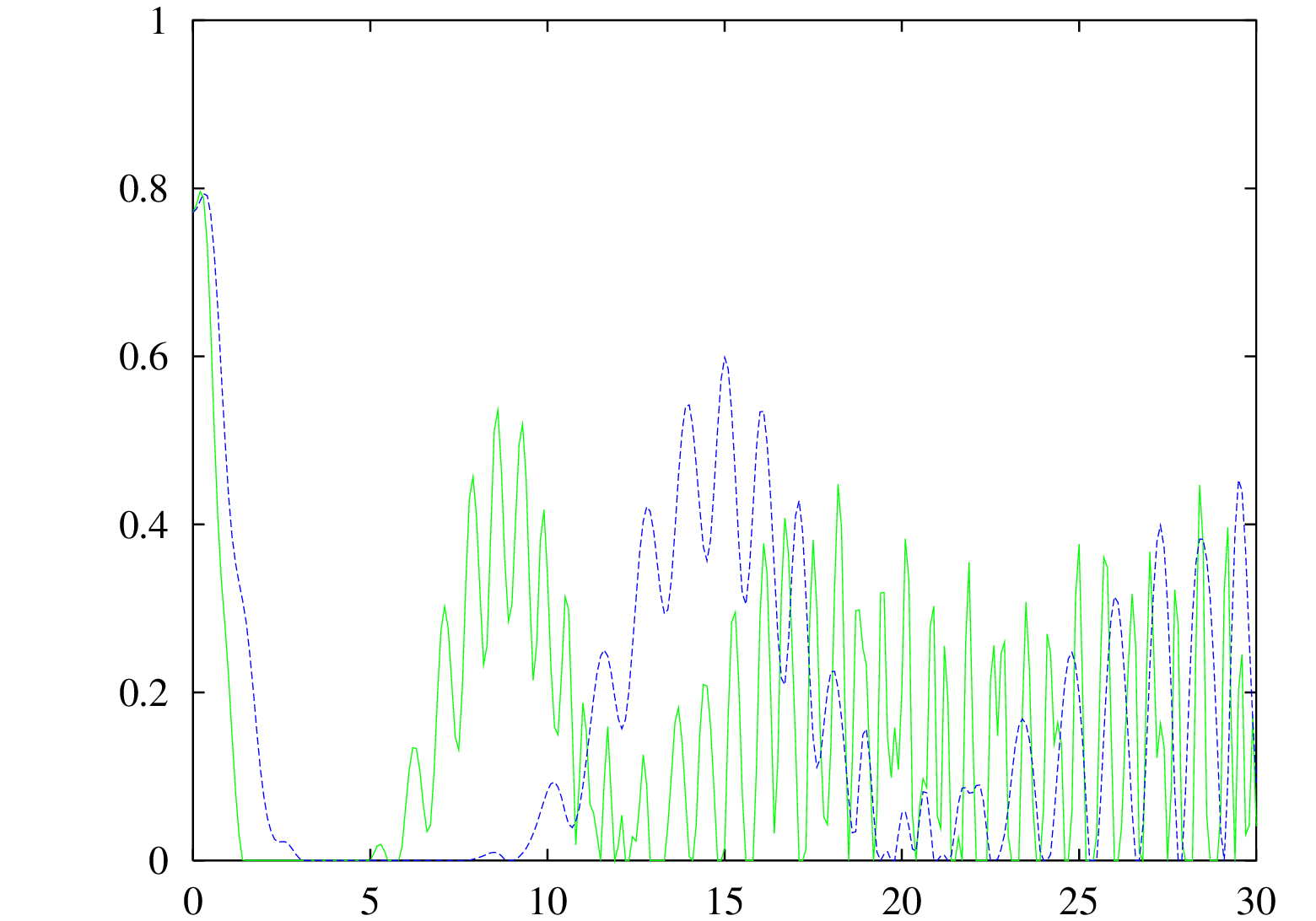}\ \put(-26,112){(a)}\
\put(-125,-10){ Scaled time}\put(-8,70){DOE} %
\includegraphics[width=20pc,height=12pc]{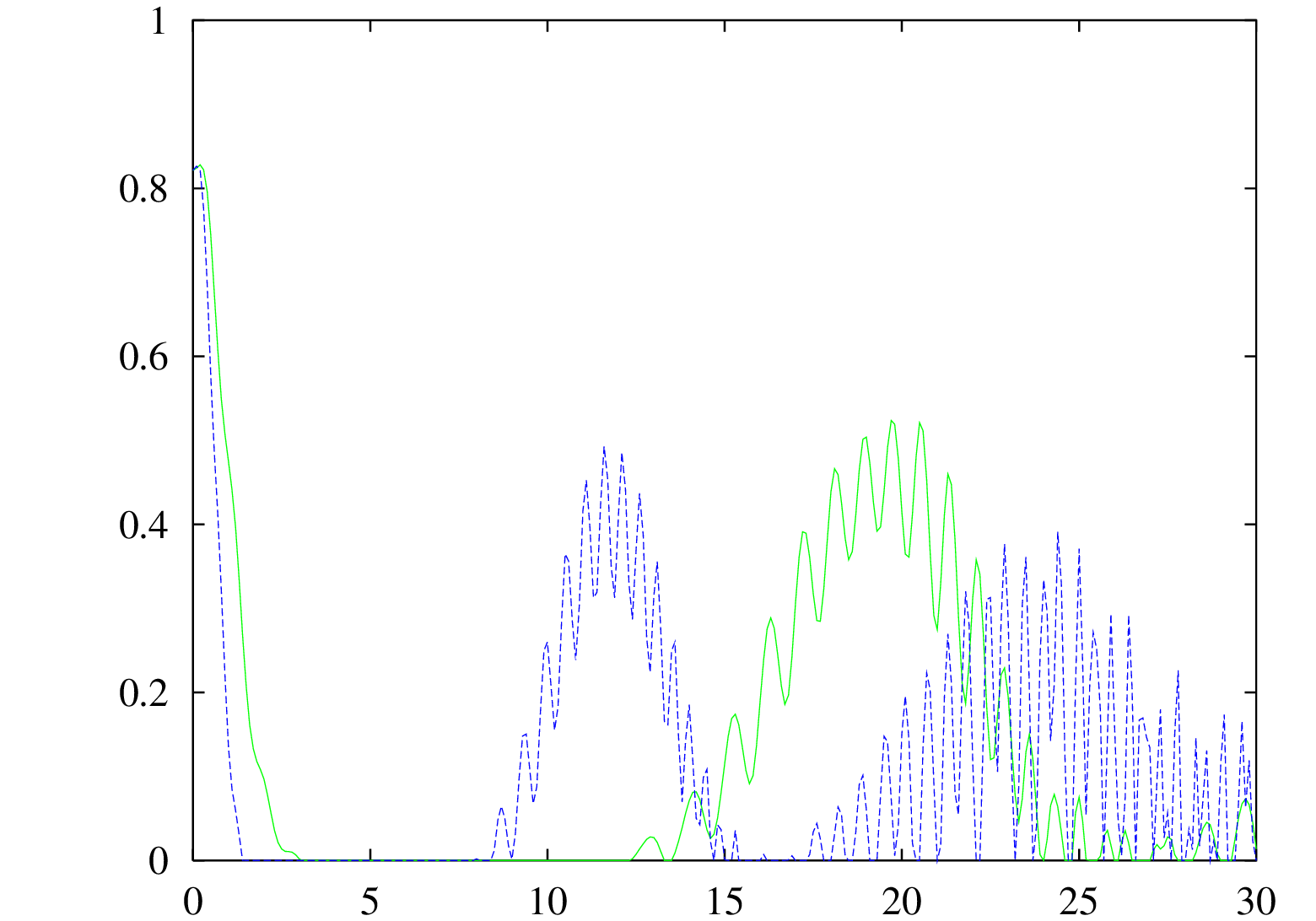}\ \put(-32,112){(b)}\
\put(-125,-10){ Scaled time}\put(-490,70){DOE}\  %
\includegraphics[width=25pc,height=10pc]{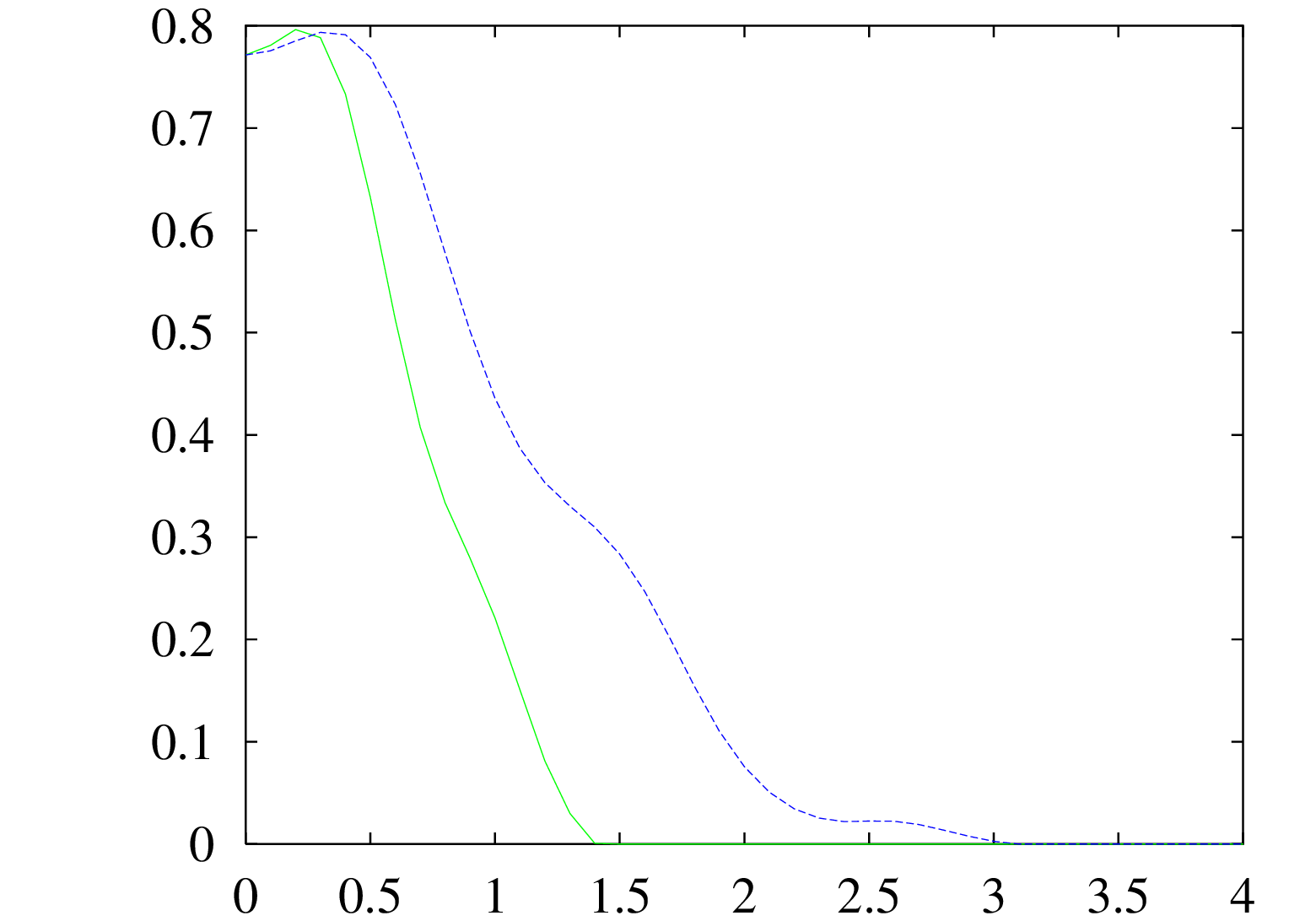} \put(-35,100){(c)}
\put(-160,-10){ Scaled time}\put(-300,65){DOE}
\end{center}
\caption{ The same as Fig.(3), but the system is prepared in the partially
entangled stat}
\end{figure}
For this reason we have plotted figures (4a) and (4b) for the mean photon
number $\bar{n}=5$ and $\bar{n}=10$, respectively. In Fig.(4a) one can see
that the entanglement reaches its maximum after onset of the interaction for
$r=0.1$ (dash line) and $r=0.8$ (solid line) and then decreases its value to
reach the minimum ($zero$ value) where the correlation between the atoms
disappeared. However, for the case $r=0.8$ we can see that after certain
period of time a weak entanglement starts to appear followed with period of
increasing in its value beside irregular fluctuations. This means that the
atoms start to gain and exchange the energy until the entanglement reaches
its maximum at $\sim 0.57.$ As soon as the entanglement reaches its maximum,
the atoms start again to loss part of their energy and the amount of
entanglement begins to decrease until to reach the minimum at $zero$. After
this period of entanglement between the atoms, the system shows fluctuations
for the rest of the considered time which reflects the exchang of the energy
between the atoms and the field. Similar behavior can be reported for $r=0.1$%
, where the entanglement in this case occurred at later time to that of $%
r=0.8,$ see Fig.(\textrm{4a}). Increasing the value of the mean photon
number $\bar{n}=10,$ one can easily realize that the maximum value for $%
r=0.1 $ and $r=0.8$ occurs after onset of the interaction as in the case $%
\bar{n}=5 $ with the same value, see Fig.(\textrm{4b}). We can also report
that the long period of entanglement between the atoms occurs first in the
case $r=0.1.$ Furthermore, the general behavior for both cases $r=0.1$ and $%
r=0.8$ are nearly the same but with fluctuations less than the case in which
$\bar{n}=5.$ It should be noted that in this case there is some delay in the
second period of the entanglement between the atoms compared with the
previous case and the period of entanglement for $r=0.8$ is greater than the
period for $r=0.1.$ Comparing these results with that obtained for the
identical atoms \cite{metwally}, we can see that the generated state
contained in the identical atoms is much larger than that obtained in the
non-identical atoms. This is due to the equal possibilities of interaction
of the two atoms together.

In Fig.(4c), we plot the degree of entanglement for small scale of time,
where we assume the same case which shown in Fig.(4a). From this figure we
see that at the zero value of the scaled time, $\lambda_1 t$, the degree of
entanglement is $\simeq0.78$, which represents the degree of entanglement of
the initial state. As the interaction goes on, the degree of entanglement is
slightly increased and then decays as the time increases. The sudden death
of entanglement \cite{Eberly} is seen clearly for small values of the
coupling parameter($r=0.1$). For large values of the coupling parameter ($%
r=0.8$),the degree of entanglement decrease smoothly. So one can say that
the survival time of entanglement increases as one increases the the
coupling constant.

\section{Impurity and the dynamics of the information}

\begin{figure}[tbp]
\begin{center}
\includegraphics[width=18pc,height=12pc]{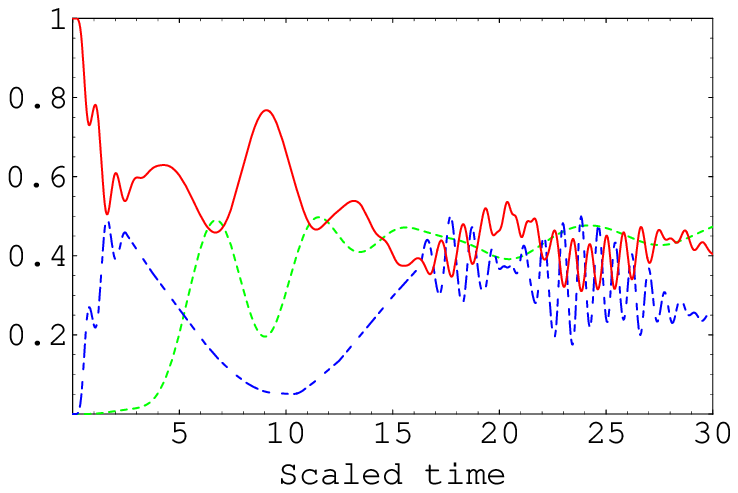}\ \put(-30,30){(a)}
\put(-200,80){$\xi$} \includegraphics[width=18pc,height=12pc]{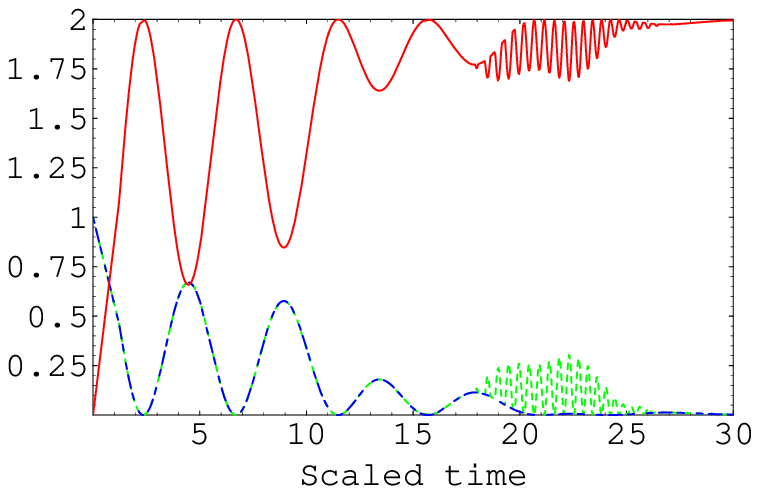}\
\put(-30,30){(b)} \put(-215,80){$I_{l\&n}$}
\end{center}
\caption{ (a) The impurity $\protect\xi$, for the density operators $\protect%
\rho_1$ (dot curve), $\protect\rho_2$(dash-dot curve) and $\protect\rho_{12}$%
( solid curve) with $R=0.1$ and $\bar n=10$ (b) The non-local information, $%
I_{non-Local}$ (for short $I_{n}$) for $\protect\rho_{12}$ (solid curve),
the dot and dash dot curve are the local information, $I_{Local}$( for short
$I_l$) for $\protect\rho_1$ and $\protect\rho_2$respectively, where $R=0.8$
and $\bar n=10$.}
\end{figure}
Our target in this section is to quantify the amount of the local and
non-local information for the present system. To reach this goal,\ let us
first define the local and non-local information which can be defined from
the following situation: Suppose we have a source supplies two users Alice
and Bob with qubit (atoms) to code their information. In this case we can
say that these information are a local information. On the other hand, if we
assume that the atoms are forced to pass through a cavity, then the two
atoms will entangled with each other and then interact with the cavity
field. As a resultant of this interaction the local information will be
transferred between the two atoms and whence the atomic system will behaves
as an entangled state or as a separable state. In this case the information
is called non-local information. Since the main purpose of the users is to
distill as much as possible from the information. Therefore to reach this
target one may seek to quantify the amount of the local information. For
this reason let us use the measure introduced in reference \cite{20} to
quantify such information which is defined by%
\begin{equation}
I_{Local}=\left( 2F_{0}(\rho )-1\right) ^{2}  \label{17}
\end{equation}%
where
\begin{equation}
F_{0}(\rho )=\max_{\mathcal{A}\in SU(2)}\langle \phi |\mathcal{A}\rho
\mathcal{A}^{\dagger }|\phi \rangle ,  \label{18}
\end{equation}%
and $|\phi \rangle $ is the initial state of the atomic system, while $%
\mathcal{A}$ is the unitary operator used by Alice and Bob to maximize their
local information. The total local information for the two qubit is given by
\begin{equation}
I_{Local}^{t}=\sum_{j=1}^{2}I_{Local}^{j},  \label{19}
\end{equation}%
and the non-local information is defined such as
\begin{equation}
I_{non-Local}=2-\sum_{j=1}^{2}I_{Local}^{j}  \label{20}
\end{equation}

To investigate the dynamic of the impurity for the individual states of the
qubits as well as the total state, one can use the definition
\begin{equation}
\xi _{k}=1-tr\{\rho _{k}^{2}\},\qquad k=1,2,12  \label{21}
\end{equation}%
This quantity $\xi_{k}$, measures the distance between the given state and
its purity, for more details see \cite{21,22}. Using similar parameters as
in the previous section we have plotted the dynamics of the impurity, the
local and non-local information for the individual qubits and the total
state. For example, in figures (5a,b) we have considered the case in which
the atomic system is prepared in the ground state with fixed value of the
photon number $\bar{n}=10$, while the ratio of the coupling parameter $r=0.1$%
. From these figures one can see that after onset of the interaction the
impurity for one individual qubit $\rho _{2}$ increases while it decreases
for the total state $\rho _{12}$ so that both of them approaches each other
without intersection. In the meantime the impurity of the qubit $\rho _{1}$
starts to increase, however, after short period of time to intersect with
the individual quibt $\rho _{2}$ as well as with the total state $\rho _{12}$%
. This is followed by certain periods of time where similar behavior can be
seen between the total states $\rho _{12}$ and individual qubit $\rho _{1}$,
where the amount of increment in $\rho _{12}$ is nearly equal to the amount
of decrement in $\rho _{1}.$ This behavior disappeared at later time where
one can see irregular fluctuations with interference between the impurity of
individual qubits $\rho _{1}$, $\rho _{2}$ and the total state $\rho _{12}$,
see Fig.(5a). This means that the information between each qubits as well as
between the total states are lost.

On the other hand we have plotted Fig.(5b) to display the dynamics of the
local and non-local information keeping all parameters unchanged. In this
case it is also seen that at $t=0$, the local information for both qubit is
maximum, while for the total state is zero. However, the individual qubits $%
\rho _{1}$ and $\rho _{2}$ decrease their values after onset of the
interaction which leads to increase the value of the total state $\rho
_{12}. $ Furthermore, the individual qubits show simultaneously fluctuations
with decreasing in their value as the time increased. However, after
considerable period of time one can see disturbance in the local information
for the individual qubit $\rho _{1}$ corresponding to irregular rapid
fluctuations in the non-local information. This means that the non-local
information would increase its minimum as soon as local information start to
die out, see Fig.(5b).

Different behavior can be observed for impurity as well as local and
non-local information when we increase the ratio of the coupling parameter $%
r=0.8$. For example, after short period of time one can see from figure (5c)
that a considerable reduction in the total state $\rho _{12}$ to reach value
below $0.4,$ this is corresponding to an increase in the impurity for each
qubits $\rho _{1}$ and $\rho _{2}$ individually. This emphasis on the fact
that the impurity of the total state is sensitive to the variation of the
dynamics of the impurity of each qubit. Also it is noted that the impurity
behavior of both qubits is almost the same up to the time $t\sim 12.$ This
in fact is due to the closed value of the coupling parameters to each other.
At later time we have realized that there is an interference between the
impurity for the qubits and the total state. This would leads to an increase
in the noise of the system and consequently reduction in the amount of the
information received. This phenomenon is also observed for $r=0.1,$ but with
more reduction in the value of \ the impurity of each individual qubits.

In figure (5d) we have plotted the local and non-local information for $%
r=0.8 $ and $\bar{n}=10,$ where we can see drastic reduction for the
non-local information value corresponding to increasing in the local
information value for each qubits. This is followed by a long period of time
where the local information keeps its value at minimum, whereas the
non-local information at maximum. This means that there is no transmission
for any information during this period. At later time one of the qubits
starts to show increasing in its value followed by the second qubit. In this
case it is easy to observe that the value of the non-local information is
decreased. Thus we may conclude that increasing the value of the coupling
parameter ratio leads to reduction and delay in transmitting the information.

Now let us turn our attention to discuss the effect of the mean photon
number when the system is initially in the ground state. To do so we have
plotted figures (6) taken into consideration $r=0.1$ and examined two
different cases $\bar{n}=5$ and $\bar{n}=7$. For these two cases a similar
behavior to that displayed in Fig.(5a) is seen in figures (6a,c). In the
meantime, it is noted that decreasing the number of photons leads to more
irregular fluctuations within certain period of time, more precisely between
$t=3$ and $t=20$. Moreover, less interference between the impurity of the
individual qubits and the total state can be reported in these cases, see
Figs.(6a,c). This means that to reduce the noise in the present quantum
system and to improve the quality of the transmission we have to decrease
the mean photon number. Also for the same values of the mean photon numbers $%
\bar{n}=5$ and $\bar{n}=7,$ we have considered the local and non-local
information. Simple comparison between the case in which $\bar{n}=10$ and
the cases where $\bar{n}=5$ and $\bar{n}=7 $, we realize that the general
behavior is the same. However, the main difference between them is occurred
in the non-local information, see Figs.(5b) and (6b,d). For example, at $%
t>15,$ one can realize that during the period of the irregular fluctuations
some revivals can be seen as we decrease the mean photon number.
Furthermore, at $t\approx 20$ the non-local information decreases its value
as we decrease the mean photon number.
\begin{figure}[t]
\begin{center}
\includegraphics[width=18pc,height=12pc]{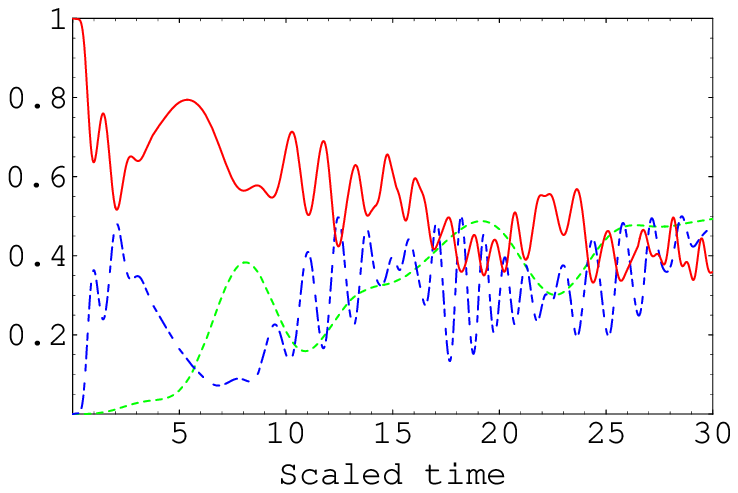}\ \put(-30,30){(a)}
\put(-200,80){$\xi$} \includegraphics[width=18pc,height=12pc]{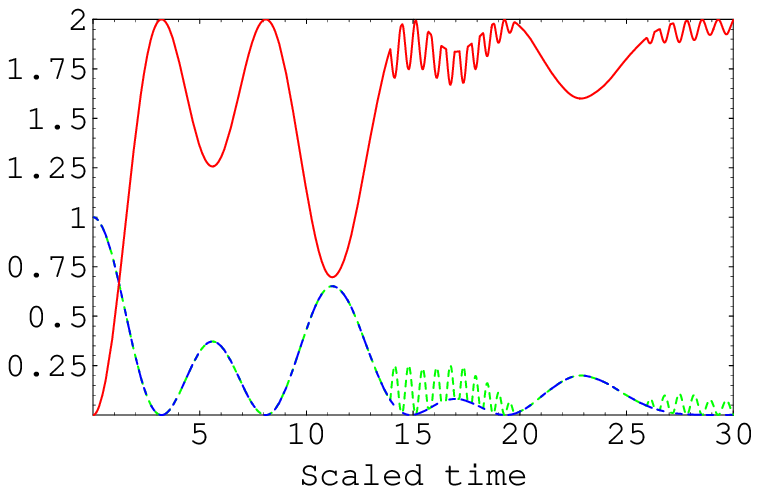}
\put(-30,30){(b)} \put(-215,80){$I_{l\&n}$}\ %
\includegraphics[width=18pc,height=12pc]{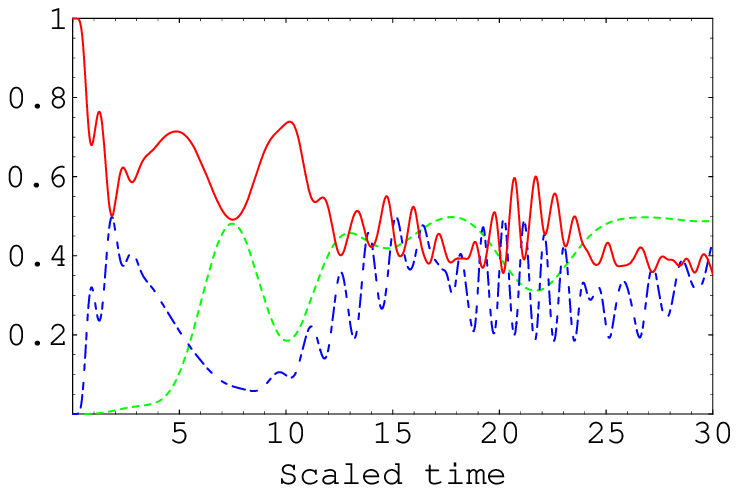} \put(-30,30){(c)}
\put(-200,80){$\xi$} \includegraphics[width=18pc,height=12pc]{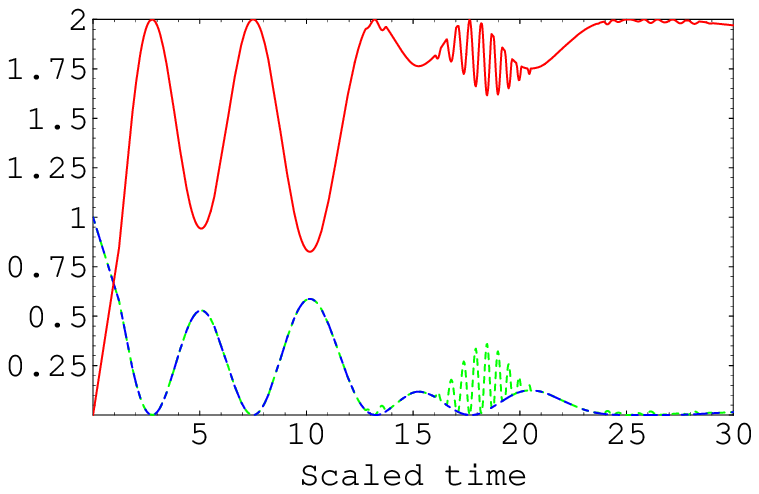}\
\put(-30,30){(d)} \put(-215,80){$I_{l\&n}$}\
\end{center}
\caption{ (a) The impurity $\protect\xi$ for the density operators $\protect%
\rho_1$ (dot curve), $\protect\rho_2$(dash-dot curve) and $\protect\rho_{12}$%
( solid curve) with $R=0.1$ and $\bar n=10$ (b) The non-local information $%
I_{n}$ for $\protect\rho_{12}$ (solid curve), the dot and dash dot curve are
the local information $I_{l}$ for $\protect\rho_1$ and $\protect\rho_2$%
respectively, where $R=0.8$ and $\bar n=10$.}
\end{figure}

Finally, let us concentrate on discussing the variation when the system is
initially in the intermediate state. For this reason we have plotted figure
(7a) to display the behavior of the impurity for $\bar{n}=7$ and $r=0.1.$ In
this case we note that after the onset of the interaction the total state $%
\rho _{12}$ decreases while the impurity of the individual qubits $\rho _{1}$
and $\rho _{2}$ increases. Also it is easy to observe that the increment in $%
\rho _{2}$ is faster than that the increment in $\rho _{1}.$ However, as the
time goes on the impurity of $\rho _{2}$ backs to decrease its value while
the impurity of $\rho _{1}$ continue its increasing. This means that the
individual qubits $\rho _{2}$ acquired part of the energy less but faster
than the energy acquired by the individual qubits $\rho _{1}$. At later time
we can see irregular fluctuations in both $\rho _{1}$ and $\rho _{2}$ as
well as in $\rho _{12}$ which indicates the appearance of some noise in the
transmission.

To analyze the local and non-local information we have plotted Fig.(7b).
Here and after onset of the interaction we see a rapid fluctuations in a
short period of time where the interference between local and non-local in
formation is pronounced. This is followed by a period of time where we see
decreasing in the local information corresponding to increasing in the
non-local information. At $t>15$ the local information increases and the
non-local decreases refereeing to better period of transmission.
\begin{figure}[tbp]
\begin{center}
\includegraphics[width=18pc,height=13pc]{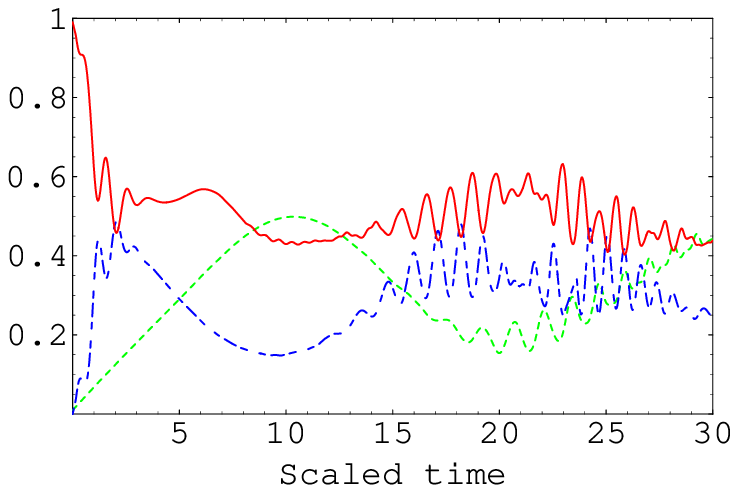}\ \put(-180,140){(a)}
\put(-200,80){$\xi$} \includegraphics[width=18pc,height=13pc]{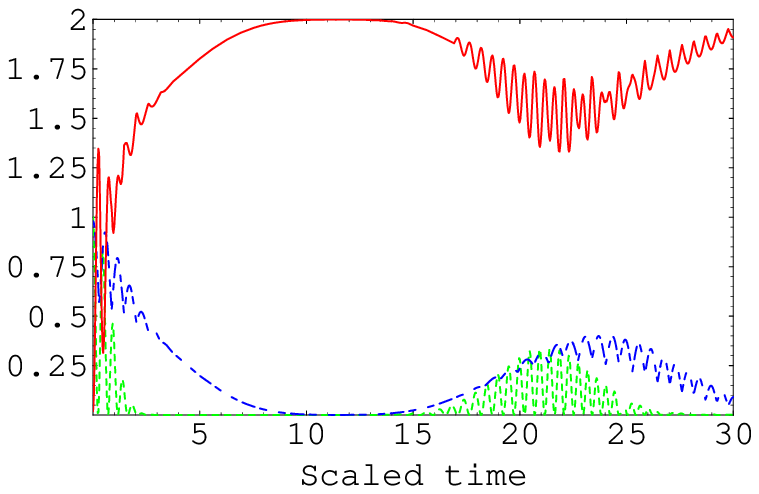}\
\put(-180,140){(b)} \put(-215,80){$I_{l\&n}$}\
\end{center}
\caption{ The same as Fig.(6), but the system is prepared in the partially
entangled state}
\end{figure}

\section{Conclusion}

In this contribution, we investigate the time evolution of atomic system
interacting with a single cavity mode from the separability point of view.
The effect of the coupling constant, between the cavity mode and the two
non-identical atoms, and the mean photon number are investigated on the
separable and entangled behavior of the atomic system. We find as one
increases the values of the coupling constant, the possibility of generating
entangled state decreases and the amount of entanglement decreases. Also, as
the mean photon number increases, the entangled intervals shifted also the
degree of entanglement decreases.

The dynamics of the impurity and the information is investigated for
different values of the coupling and initial state of the two atoms. It is
found that as the impurity increases for one qubit, decreases for the other
one. Also, the transfer of the local information into non-local depends on
the impurity of the individual qubits. If the impurity of one qubit is
maximum, then the total information is converted to be non-local. This
phenomena is too important in quantum communication as we shall see in our
next work.

\textbf{Acknowledgement:}

One of us (M.S.A.) is grateful for the financial support from the project
Math 2005/32 of the Research Center, College of Science, King Saud
University.

\end{document}